\documentclass[aps,prl,reprint,superscriptaddress]{revtex4-1}

\usepackage{graphicx}
\usepackage{calligra}
\usepackage{upgreek}
\usepackage{color}
\usepackage{inputenc,caption}
\usepackage{stackrel}
\usepackage[usenames,dvipsnames]{xcolor}
\usepackage{subcaption}
\usepackage{placeins}

\DeclareMathAlphabet{\mathpzc}{OT1}{pzc}{m}{it}

\newcommand{\be}{\begin{equation}}
\newcommand{\ee}{\end{equation}}
\newcommand{\bea}{\begin{eqnarray}}
\newcommand{\eea}{\end{eqnarray}}
\newcommand{\lb}{\label}

\newcommand{\bu}{{\bf u}}

\newcommand{\bx}{{\bf x}}

\newcommand{\btau}{{\mbox{\boldmath $\tau$}}}

\newcommand{\grad}{{\mbox{\boldmath $\nabla$}}}
\newcommand{\bdot}{{\mbox{\boldmath $\cdot$}}}

\newcommand{\hvline}{{$\!$\textbf{-}$\!$\textbf{-}$\!$\textbf{-}$\!$\textbf{-}$\!$\textbf{-}}}

\definecolor{forestgreen}{rgb}{0.33, 0.55, 0.43}
\definecolor{paleblue}{rgb}{0.4, 0.7, 1}
\definecolor{palered}{rgb}{1, 0.48, 0.25}

\begin{document}

\setlength{\abovedisplayskip}{8pt}
\setlength{\belowdisplayskip}{8pt}

\title{Thermal noise competes with turbulent fluctuations below millimeter scales}
\author{Dmytro Bandak}
\affiliation{Department of Physics, University of Illinois at
Urbana-Champaign, Loomis Laboratory of Physics, 1110 West Green Street,
Urbana, Illinois 61801, USA}
\author{Gregory L. Eyink}
\affiliation{Department of Applied Mathematics and Statistics, and Department of Physics, 
The Johns Hopkins University, Baltimore, MD, USA,
21218}
\author{Alexei Mailybaev}
\affiliation{Instituto de Matem\'atica Pura e Aplicada - IMPA, Rio de Janeiro, Brazil}
\author{Nigel Goldenfeld}
\affiliation{Department of Physics, University of Illinois at
Urbana-Champaign, Loomis Laboratory of Physics, 1110 West Green Street,
Urbana, Illinois 61801, USA}

\begin{abstract}
Turbulent flows frequently accompany physical, chemical and biological
processes, such as mixing, two-phase flow, combustion and even foraging
by bacteria and plankton larvae, all of which are in principle subject
to thermal fluctuations already on scales of several microns. Nevertheless 
the large separation between the millimeter scale at which turbulent fluctuations 
begin to be strongly damped and the mean free path of the fluid has been 
generally assumed to imply that thermal fluctuations are irrelevant to the
turbulent dissipation range. Here we use statistical mechanical estimates to show
that thermal fluctuations are not negligible compared to turbulent
eddies in the dissipation range. Simulation of the Sabra shell model
shows that intermittent bursts of turbulence lead to a fluctuating
length scale below which thermal fluctuations are important: over three 
decades of length, from sub-millimeter scales down to the mean free path, 
thermal fluctuations coexist with hydrodynamics.  Our results imply that thermal fluctuations
cannot be neglected when modeling turbulent phenomena in the far
dissipation range.
\end{abstract}

\maketitle

Thermal effects are known to play a prominent role in many physical,
chemical and biological processes in molecular fluids. These include high
Schmidt/Prandtl-number scalar mixing \cite{donev2014reversible}, droplet and bubble
formation \cite{chaudhri2014modeling, gallo2020nucleation}, locomotion of micro-organisms
\cite{gotze2010mesoscale}, combustion \cite{lemarchand2004fluctuation, bhattacharjee2015fluctuating},
and others. Often the relevant flows are turbulent, in which case all current approaches to model
these processes --- high Schmidt-number scalar mixing \cite{clay2018gpu, buaria2020turbulence},
droplet and bubble formation \cite{elghobashi2019direct, milan2020sub}, cellular motility
\cite{durham2013turbulence, wheeler2019not}, and combustion
\cite{donzis2010batchelor, sreenivasan2004possible, driscoll2008turbulent, echekki2010turbulent}
--- focus on turbulent fluctuations due to fluid
inertia damped by viscosity and ignore thermal effects completely. This
neglect is commonly justified by the idea that there is a strong
separation between hydrodynamic scales dominated by turbulent
fluctuations and extremely small scales of order the mean-free path
length where thermal fluctuations begin to play a role
\cite{vonneumann1963recent}. The smallest
turbulent fluid scales below which eddies are damped by viscosity,
known as the {\it dissipation range}, typically occurs at millimeter
scales and there is currently intensive effort, by computation
\cite{khurshid2018energy, gorbunova2020analysis,
buaria2020dissipation}, theory
\cite{pauls2020analytic,canet2017spatiotemporal,gibbon2021correspondence}
and experiment
\cite{debue2018experimental,gorbunova2020analysis,debue2021three}, to
understand the dynamics and statistics at these scales. The motivation
ranges from unraveling the nature of turbulence itself, and the
associated unsolved issue of formation of singularities
\cite{fefferman2006existence}, to the phenomena and practical
applications mentioned above and others.

The basic underlying assumption in turbulent flows of a viscous dissipation range
well-separated from thermal fluctuations is at first sight surprising. This seemingly contradicts the {\it fluctuation-dissipation theorem} of statistical physics (e.g. see \cite{dezarate2006hydrodynamic}), which
implies that thermal fluctuations and dissipation should be
intrinsically tied together, even in a far from equilibrium system
where local thermal equilibrium can be justified.  The neglect of
thermal effects in the dissipation range is convenient; it allows one 
to use the Navier-Stokes partial differential equations exclusively 
to model the phenomena of interest and even this
requires a computational {\it tour de force\/} for turbulent flows
\cite{khurshid2018energy, gorbunova2020analysis,
buaria2020dissipation}. On the other hand, the hydrodynamic equations consistent with the fluctuation-dissipation
theorem for fluids locally in thermal equilibrium were formulated
already by Landau and Lifshitz in 1959 \cite{landau1959fluid,
dezarate2006hydrodynamic}, and the effects of thermal fluctuations
measured at the onset of Rayleigh-B\'enard convection 
\cite{wu1995thermally}. As far as turbulence is concerned, most prior 
discussions \cite{hosokawa1976ensemble,ruelle1979microscopic,machavcek1988role}
suggest that thermal noise plays only only a secondary  role in the selection 
and uniqueness of the stationary measures characterizing turbulence. 
To what extent are these seemingly abstract
considerations relevant to real flows, especially given the large
separation in scales between the mean free path and those where
hydrodynamic dissipation occurs?

The purpose of this Letter is to provide theoretical arguments and
numerical evidence that thermal noise effects manifest throughout the
turbulent dissipation range, in contradiction to widespread belief.  Our
results imply that a nontrivial interplay between turbulent and thermal
fluctuations must occur at small scales for nearly all flows in Nature
and in the laboratory. Our results support pioneering ideas of Betchov 
\cite{betchov1957fine,betchov1961thermal,betchov1964measure}, who reached 
similar conclusions already in 1957, and extend them to account for the 
effects of inertial-range intermittency. 

In the low Mach number ($Ma$) limit, {\it fluctuating hydrodynamics}
adds a stochastic stress term $\grad\cdot \btau$ to the incompressible
Navier-Stokes equation for the velocity $\bu$ satisfying \cite{forster1976long,forster1977large,usabiagaa2012staggered,donev2014low}
$\grad\bdot\bu=0$ and \be \partial_t\bu + (\bu\cdot\grad)\bu = -\grad p
+ \nu \Delta \bu + \grad\cdot \btau \lb{FNS}.\ee
In accord with the general fluctuation-dissipation relation, the stress
$\tau_{ij}(\bx,t)$ which models thermal fluctuations is a  Gaussian
random field with mean zero and covariance
\begin{eqnarray}
\langle \tau_{ij}(\bx,t) \tau_{kl}(\bx',t')\rangle& = &\frac{2\nu k_BT}{\rho}
\left(\delta_{ik}\delta_{jl}+\delta_{il}\delta_{jk} -\frac{2}{3}\delta_{ij}\delta_{kl}\right)\cr
&& \hspace{30pt} \times \delta^3(\bx-\bx')\delta(t-t') \lb{FDR}, \end{eqnarray}
which is proportional to the kinematic viscosity $\nu,$ the dissipative transport coefficient
of the fluid, and absolute temperature $k_BT$ in energy units set by Boltzmann's constant $k_B$.
The other parameter appearing in (\ref{FDR}) is $\rho,$ the mass density of the fluid. It is precisely Eq.(\ref{FNS})
that one must solve numerically \cite{usabiagaa2012staggered,donev2014low}, or otherwise, in order
to model molecular fluids at low Mach number.

\textit{Noise estimate.} We now rehearse standard arguments \cite{vonneumann1963recent}
that thermal effects in fluids become important only at scales comparable to
the mean free path length $\lambda_{mfp}$, and then explain why these
arguments are not valid. The velocity fluctuations at length scale
$\ell$ from thermal fluctuations can be estimated as 
\begin{equation}  u_\ell^{th}\sim c_{th}/(n\ell^{3})^{1/2} = c_{th}
(\ell_{int}/\ell)^{3/2}, \label{tvth-ell} \end{equation} 
where  $c_{th}$ is the speed of sound, 
$n$ is the number density and $\ell_{int}$ is the typical interparticle
distance. The standard argument asks when thermal fluctuations at the
scale $\ell$ are of the same order as a typical flow velocity $u$;
equating them, we find that $\ell \sim \ell_{int} (Ma)^{-2/3}$, where
$Ma=u/c_{th}$ is the corresponding Mach number. By this argument,
thermal fluctuations become important only at scales comparable to
$\ell_{int}$, which is of the same order as $\lambda_{mfp}$ in liquids
and $\ll \lambda_{mfp}$ in gases. Even for $Ma\ll 1,$ the thermal
fluctuation velocities are still much smaller than $u$ until $\ell$
becomes comparable to $\ell_{int}$, and so can be neglected.  The flaw
in this argument is that the thermal velocity at scale
$\ell$ should not be compared with the mean velocity or r.m.s. velocity
$u$ of the eddies of largest size $L$.  Instead it is more appropriate
to compare $u^{th}_\ell$ with the r.m.s velocity $u_\ell$ of turbulent eddies 
at the {\it same\/} scale $\ell$. In the dissipation range
$u_\ell$ drops exponentially with decreasing $\ell$ so that
thermal fluctuations very quickly become competitive with turbulent
fluctuations, as is shown by the following quantitative estimate.

\begin{figure}
\begin{center}
\includegraphics[width=264pt]{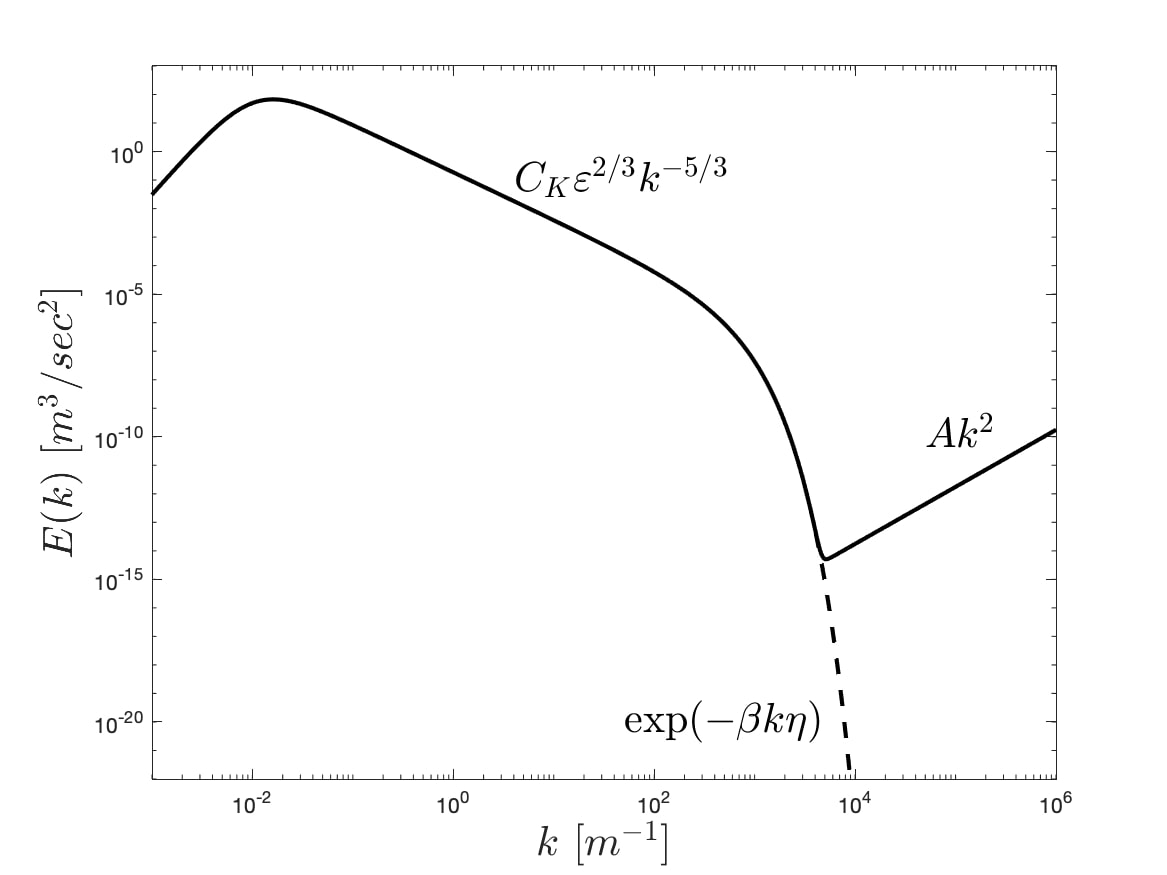}
\end{center}
\caption{Plot of the model turbulent energy spectrum (\ref{karman}) for the parameters of the atmospheric boundary
layer (\ref{ABL-param}) and for a typical Reynolds number $Re=10^7$, as a solid line. The dashed line is the spectrum
with no thermal noise.}
\label{spectrum} \end{figure}


We consider a standard model of the turbulent energy
spectrum devised by von K\'{a}rm\'{a}n \cite{vonkarman1948progress}, supplemented with an
exponential factor to represent decay of turbulent fluctuations due to viscous effects.
The estimate (\ref{tvth-ell}) of thermal velocity fluctuations can be equivalently restated as a $k^2$
equipartition energy spectrum, as first noted for equilibrium fluids in 3D by Lee \cite{lee1952some},
and Hopf \cite{hopf1952statistical}. The entire model spectrum then becomes
\be E(k)=C_K (\varepsilon L)^{2/3}\frac{L^5k^4}{(1+(kL)^2)^{17/6}}\exp(-\beta k\eta)
+ Ak^2 \lb{karman},\ee
where $C_K$ is the Kolmogorov constant, $\beta$ is the exponential decay rate in the
far dissipation range and $A=\frac{4\pi}{(2\pi)^3} \frac{k_BT}{\rho}$ is the factor
associated to thermal fluctuations in 3D. This model is plotted in Fig.\ref{spectrum}
with parameter values appropriate to the atmospheric boundary layer (ABL)
\cite{garratt1994atmospheric}
\begin{eqnarray}
&& L=10^5 \ cm,\quad \varepsilon=400 \ cm^2/sec^3, \quad \nu = 0.15\ cm^2/sec, \cr
&&\rho=1.2 \times 10^{-3}\ g/cm^3, \quad T=300^\circ K, \lb{ABL-param}
\end{eqnarray}
and with typical values of $\beta=7$ \cite{khurshid2018energy} and $C_K=1.6$
\cite{donzis2010bottleneck}. One can see in Fig.\ref{spectrum} the familiar regimes
of turbulent flow, the inertial range with $E(k) \propto k^{-5/3}$, which transitions
into the dissipation range with $E(k) \propto \exp(-\beta k\eta)$ around the Kolmogorov scale
$\eta = (\nu^3/\epsilon)^{1/4} \sim 0.5\ \mathrm{mm}$. 
However, the model (\ref{karman}) predicts that once thermal fluctuations are included,
this exponential decay is completely replaced by the $k^2$ equipartition spectrum.
The presence of this thermal component on top of the turbulent spectrum does not 
automatically follow from the inclusion of thermal fluctuations in the 
dynamical equations (\ref{FNS}), but can 
arise through the separation of time-scales, if it turns out that viscosity and thermal
noise are more dominant than the nonlinear term in (\ref{FNS}) for dissipation-range  
scales. Thus it is necessary to study this issue quantitatively, which we do below.  
We may note that thermal fluctuations have been considered already in numerical 
simulations of superfluid turbulence using 
Gross-Pitaevskii equation \cite{shukla2019quantitative}, and similar equipartition spectra observed.

We can estimate on order of magnitude the wavenumber $k_{eq}$ at which exponential decay is
overtaken by thermal equipartition just by equating the two spectra, as
$ u_\eta^2 \eta \mathrm{e}^{-k \eta} \sim \frac{k_B T}{\rho} k^2 $
where $u_\eta=(\varepsilon\nu)^{1/4}$ is the Kolmogorov velocity. This leads to a
formula for $k_{eq}$ of the form \be \lb{thetaK} (k_{eq} \eta)^2 \mathrm{e}^{k_{eq}
\eta} = 1/\theta_K \lb{keq} \ee in which appears the dimensionless
ratio \be \theta_K =\frac{k_BT}{\rho u_\eta^2 \eta^3}. \ee
The quantity $\theta_K$ is a
new dimensionless number group that characterizes the turbulent flow,
in addition to the Reynolds number $Re.$ Substituting into
(\ref{thetaK}) the values of parameters (\ref{ABL-param}) characteristic
of the ABL, we arrive at the value $\theta_K = 2.83 \times 10^{-8}.$ It
is then easy to calculate from (\ref{keq}) that $k_{eq}\eta\doteq 11,$
which in the ABL corresponds to a length $\ell_{eq}\simeq 49\ \mu$m.
In fact, due to the rapid increase of the exponential in (\ref{keq}),
this equipartition scale $\ell_{eq}$ is very insensitive to the precise
value of $\theta_K$ and always takes a value only a factor of 10 or so
smaller than $\eta.$ This result is strikingly different from the naive
expectation that thermal effects become important only at length scales
comparable to the mean free path, which is $\lambda_{mfp} = 68 \
\mathrm{nm} \sim  \eta/10^{4}$ in the ABL. Alternatively, using parameters from water experiments of \cite{debue2018experimental}, $\eta=16\ \mu\mathrm{m}$ and $\theta_K\doteq 2.5\times 10^{-7},$ we estimate that thermal 
fluctuations should be significant from just below $16\ \mu\mathrm{m}$ 
down to $\lambda_{mfp}\doteq 0.25\ \mathrm{nm}$. In summary, naturally occurring and practically relevant turbulent flows possess several decades of length scales where a hydrodynamic approximation
is justified and yet fluctuations are dominated by thermal effects.

The previous arguments and conclusions are nearly identical to those presented 
by Betchov \cite{betchov1957fine}, except that, rather than an energy spectrum 
exhibiting exponential decay, he assumed a fast power-law decay $\sim k^{-n}$ 
with $n\doteq 6$-$7$ in the far-dissipation range, as predicted by 
contemporary theories of Heisenberg \cite{heisenberg1985statistischen} 
and others. To assess these arguments and to extend them beyond the energy
spectrum to higher-order moments sensitive to intermittency effects, we now 
consider a simple dynamical model of turbulence which is believed to capture 
the nontrivial multifractal scaling properties of turbulence.

\begin{figure}
\begin{center}
\includegraphics[width=240pt]{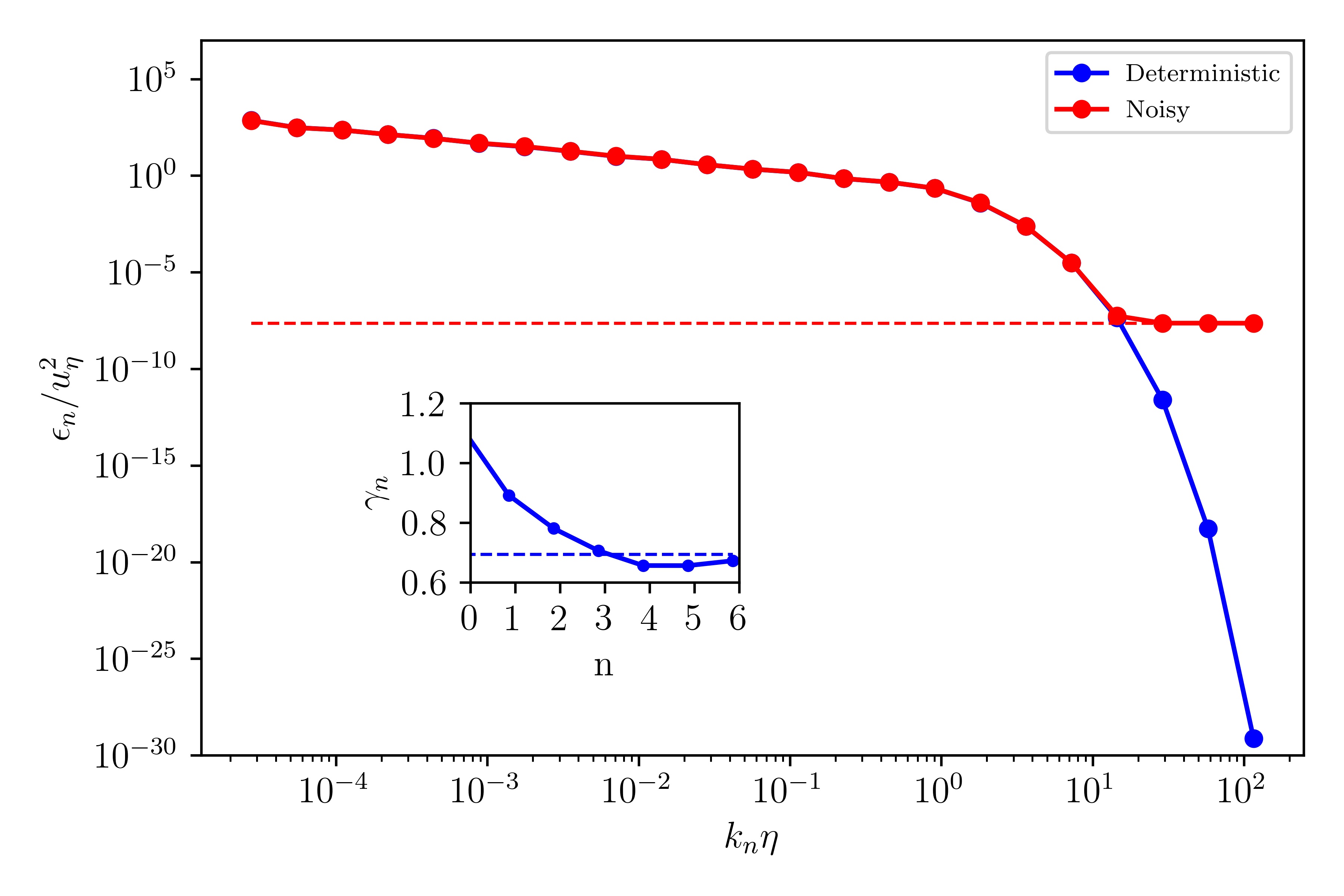}
\end{center}
\caption{The energy spectrum of the deterministic model (\textcolor{blue}{\hvline}), noisy model (\textcolor{red}{\hvline}), and the thermal equipartition value $2\theta_K$ (\textcolor{red}{-$\,$-$\,$-}). {\it Inset:} The local stretching exponents (\ref{local-stretch}) for the deterministic model spectrum 
 (\textcolor{blue}{{\bf $\bullet$}}) and its theoretical value
 in (\ref{stretch-decay}) (\textcolor{blue}{-$\,$-$\,$-}). }
\lb{comp-spectrum}  \end{figure} 


\textit{Numerical simulations.} To further understand the interplay of
turbulent and thermal fluctuations we turn to numerical simulations.
Unfortunately, direct numerical simulations of the fluctuating
Navier-Stokes equation (\ref{FNS}) with sufficient range of scales to
study how rare intense events at high Reynolds number interact with
thermal fluctuations are not yet feasible, even with the largest
supercomputers in the world. We therefore resort to shell models,
simplified dynamical systems that are widely used as surrogates of
Navier-Stokes equations in order to perform high Reynolds number
simulations \cite{biferale2003shell} of scaling properties. In
particular, we employ the Sabra shell model \cite{lvov998improved} supplemented with stochastic terms controlled by ''temperature`` to represent thermal noise. It is defined by a set of coupled
nonlinear stochastic ODEs that describe evolution of complex variables
$u_n$ defined at discrete wavenumbers $k_n= k_0 2^n,$ $n=0,1,2,...,N$. Here $u_n$ represents the
``velocity'' of an eddy of size $1/k_n$. At large times the Sabra model with fixed forcing
attains a statistical steady-state with constant flux of energy $E =
\sum_n \epsilon_n = \frac{1}{2} \sum_n |u_n|^2 $ due to the cascade across
scales in the inertial range. To understand the relative importance of
turbulent and thermal fluctuations, we simulated the system for times
large enough to represent the steady-state with the parameters in
(\ref{ABL-param}) for the ABL, achieving the Reynolds number $Re\sim 10^6$. The simulation was done twice, once with thermal noise, once without. Further details on stochastic Sabra model and our numerical stochastic integration scheme, as well as convergence tests can be found in Supplementary information and elsewhere \cite{eyink2021dissipation}.

\textit{Discussion of results.} The numerical results confirm our basic
claim that thermal equipartition is achieved at scales close to
$\eta/10$. This can be readily seen in Fig. \ref{comp-spectrum}, which
depicts the mean energy spectrum $\bar{\epsilon}_n = 1/2 \langle
|u_n|^2\rangle$ of the deterministic and noisy models. Note that shell
models describe fluids effectively in $0$ space dimensions, so that
equipartition corresponds to a constant value $(1/2)\varrho\langle
|u_n|^2\rangle=k_B T,$ or $1/2 \langle |u_n|^2\rangle=\theta_K$ in
dimensionless Kolmogorov units. The deterministic model in the
dissipation range displays stretched-exponential decay \be \langle
\epsilon_n \rangle \sim \exp(-c(k_n\ell_K)^\gamma), \quad
\gamma=\log_2\left(\frac{1+\sqrt{5}}{2}\right) \lb{stretch-decay} \ee
which was verified by plotting local stretching exponents \be \gamma_n
= \log_2 \left|\ln\langle|u_{n+1}|^2\rangle\right|- \log_2
\left|\ln\langle|u_{n}|^2\rangle\right| \lb{local-stretch} \ee (see
Inset of Fig. \ref{comp-spectrum}). Such decay is expected for
deterministic shell models
\cite{schorghofer1995viscous,lvov1998universal}, but is not observed at
all once thermal noise is included. The stretched exponential decay is
then completely erased at shell-numbers above an equilibration scale
$N_e=4$, where the modal energy is close to the equipartition value
$\theta_K$. This equilibration wavenumber is just 16 times larger than
the Kolmogorov wavenumber. It should be emphasized that our shell model
must underestimate how rapidly the dissipation range is overtaken by
thermal noise in comparison to true hydrodynamic turbulence. Indeed,
the latter has a faster exponential spectral decay in the dissipation
range \cite{khurshid2018energy} compared to stretched-exponential for
the shell model and, furthermore, the equipartition spectrum $E(k)$ is
rising in hydrodynamic turbulence $\propto k^2$ and is decaying as
$E(k_n):=\bar{\epsilon}_n/k_n\sim \theta_K/k_n$ in the shell model.

\begin{figure}[h!]
\begin{center}
\includegraphics[width=240pt]{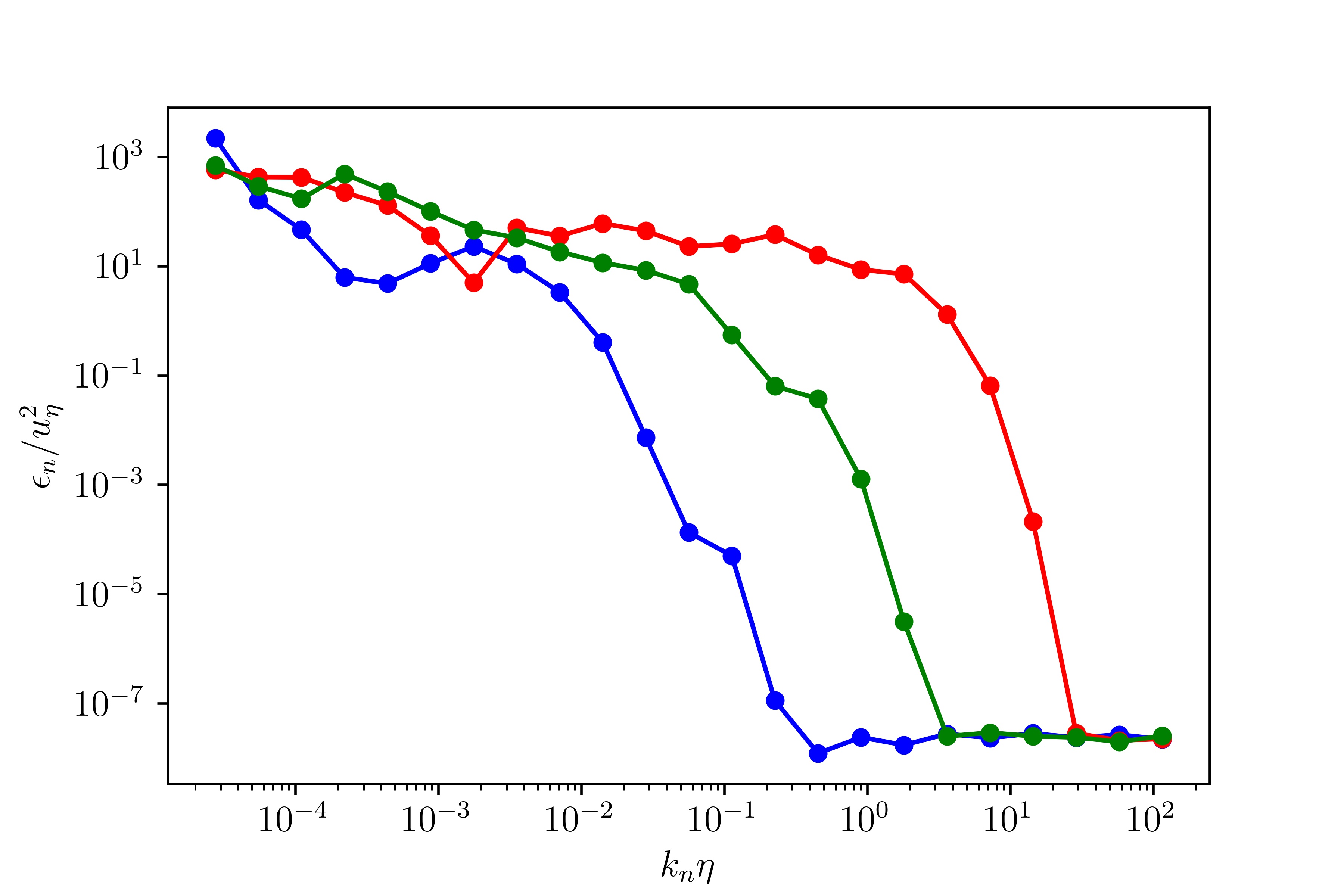}
\end{center}
\caption{Modal energies of a lull (\textcolor{blue}{{\bf blue}}), 
a burst (\textcolor{red}{{\bf red}}) and a typical realization
(\textcolor{forestgreen}{{\bf green}}), 
averaged over 1 Kolmogorov time, plotted versus $n$.}
\label{bursts} \end{figure}

In contrast to the dissipation range spectrum for the noisy model, the
inertial range energy spectrum shown in Fig.~\ref{comp-spectrum} is
seemingly unchanged by thermal noise, which is an indication that the
deterministic model is adequate for modeling fluctuations in the
inertial range. The latter is known to exhibit intermittency realized
through bursts that form at large scales and propagate to the
dissipation range. Such bursts should be expected to cause the
equilibration shell-number $N_e$ to fluctuate strongly in time. To
investigate this issue, we defined an instantaneous equilibration
shell-number $N_e(u)$ as the smallest integer such that time-averages
of $\epsilon_n$ over one Kolmogorov time are below $2 \theta_K$ for all $n \geq N_e(u)$. We illustrate
this concept in Fig. \ref{bursts}, which plots the modal energy
averaged over one Kolmogorov time as a function of wavenumber, for one
typical shell-model realization and two extreme ones. We interpret the
two extreme events as a large burst pushing $N_e(u)$ to a high value
and as a deep lull allowing $N_e(u)$ to recede to a lower value. The
underlying picture of intermittency is accepted for turbulence both in
Navier-Stokes equation and in shell models, where it is attributed to
self-similar solutions \cite{PhysRevE.62.3592, PhysRevE.87.053011} that
in the inviscid limit propagate in finite time to infinite wavenumbers.
The same mechanism underlies the fluctuating nature of the ``local
viscous shellnumber'' in the deterministic model \be
N_{vis}(u):=\min\left\{n:\, \frac{|u_n|}{\nu k_n}\leq 1\right\},
\lb{Nvis} \ee defined in analogy to the ``local viscous/cutoff length''
$\eta(\bx,t) \sim \frac{\nu}{\delta_\eta u(\bx,t)}$ of hydrodynamic
turbulence \cite{paladin1987degrees,schumacher2007sub}.

\begin{figure}
\begin{center}
\includegraphics[width=240pt]{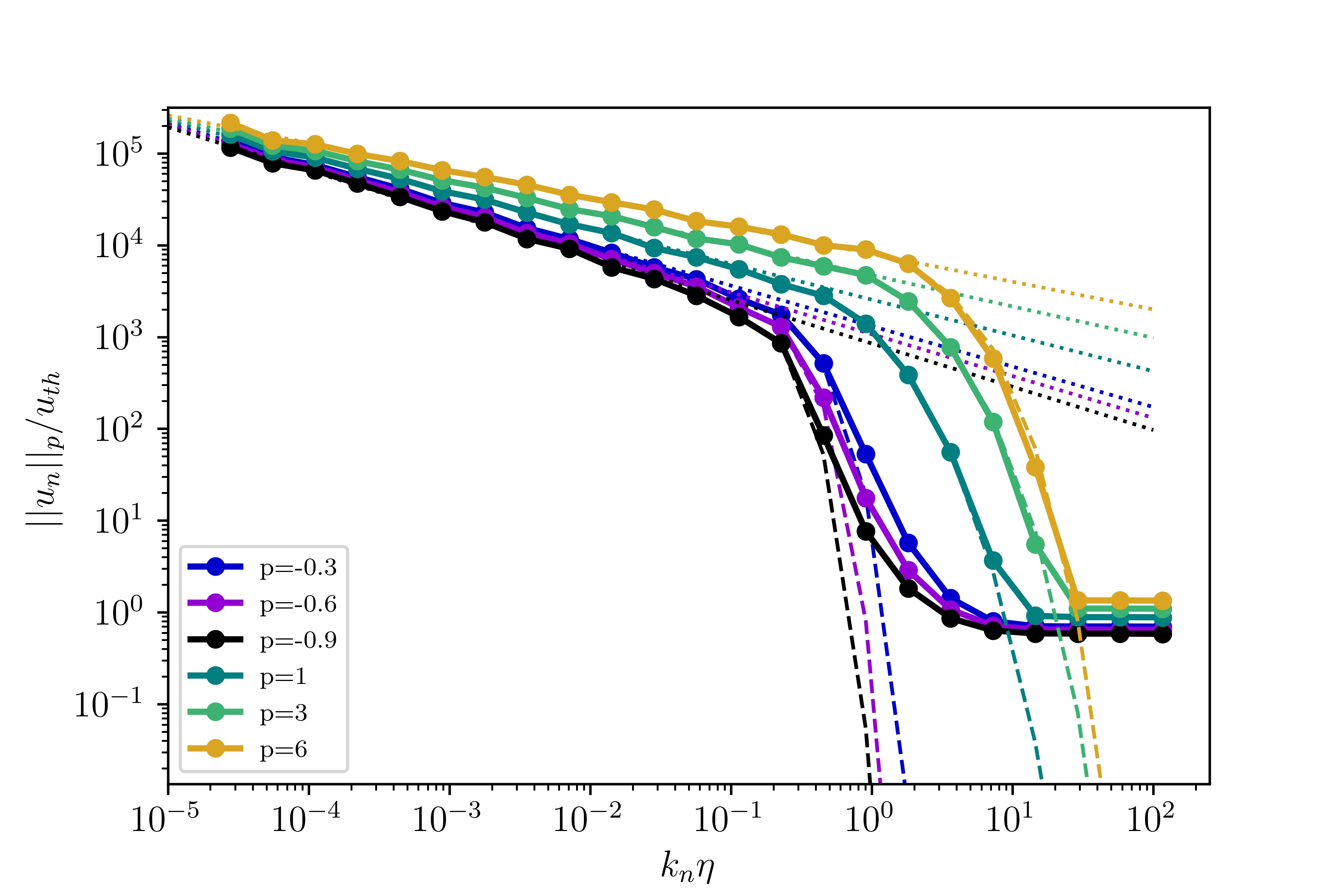}
\end{center}
\caption{Structure functions for the deterministic model (dashed lines), for the noisy model (solid lines), and power-law fits in the inertial range (dotted lines).}
\label{structfun} \end{figure}

To characterize the statistical properties of the equilibration
$N_e(u)$ and viscous $N_{vis}(u)$ scales we plot the probability
density function (PDF) of their temporal fluctuations for the noisy Sabra model
in Fig. \ref{multi-pdf}. The PDF of $N_e(u)$ is well approximated at
the core of the distribution by a discrete Gaussian
$p(n)=e^{-(n-2)^2/2}/\Theta$ \cite{agostini2019discrete}. This is
somewhat surprising given the obvious asymmetry between small and large
scales, but presumably such asymmetry is evidenced better in the
extreme tails of the PDF which are not well-resolved with our data from
300 turnover times. Plotted as well in Fig. \ref{multi-pdf} is the PDF
of $N_{vis}(u),$ defined as in (\ref{Nvis}). This PDF is plotted
together with a standard multifractal model of the type developed for
deterministic Navier-Stokes \cite{biferale2008note}, here constructed
using the ansatz \cite{bowman2006links} \be N_{vis}(h)={\rm
Round}\left[\log_2(Re)\left(\frac{1}{1+h}-\frac{3}{4}\right)\right],
\lb{Nvis-MF} \ee where ``Round'' denotes rounding to the nearest
integer, and the PDF of the H\"older exponent $h$ is taken to be
$P(h)\propto Re^{D(h)/(1+h)}$ for multifractal dimension spectrum
$D(h)=\min_p\{ph-\zeta(p)\}$ with $\zeta(p)$ evaluated from our
numerical simulation (see Fig. \ref{structfun}). The multifractal model
is in reasonable agreement with the PDF of $N_{visc}$. Furthermore, for
each realization that we used to calculate the PDF's of $N_{visc}(u)$
and $N_e(u)$ we calculated a shift $\Delta N(u) = N_{visc}(u) - N_e(u)$
and its probability distribution in the steady-state. This distribution
is peaked at $\Delta n = 3$ and indeed supports an interpretation of
the fluctuations of both $N_{visc}$ and $N_e$ being due to bursting
nature of the inertial range. A reasonable interpretation is that, in
each realization, the fluctuating viscous scale is followed by a
stretched exponential decay, which gets overtaken within just a couple
of shells by equipartition at the level of thermal energy $\theta_K$.

\begin{figure}[h!]
\begin{center}
\includegraphics[width=260pt]{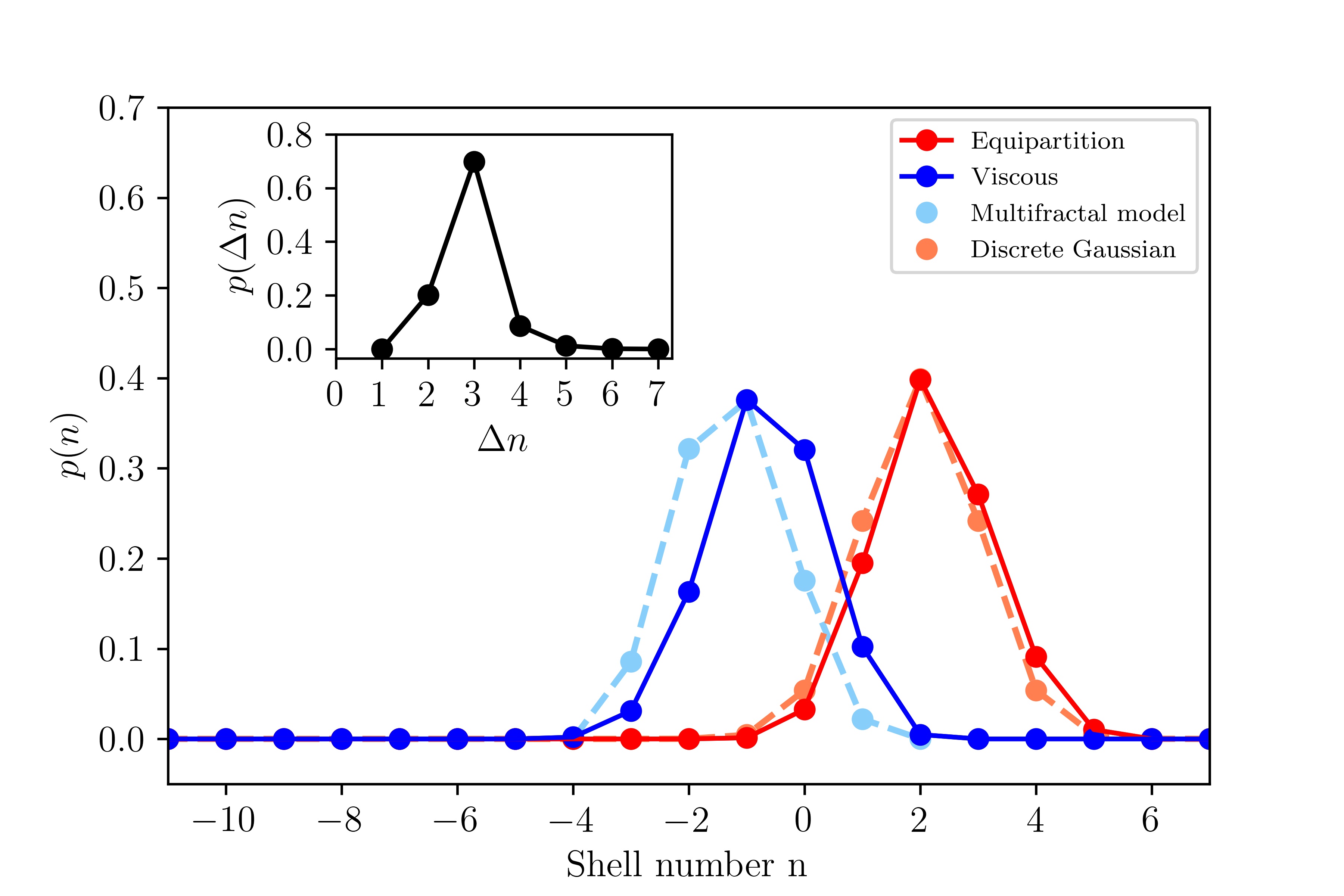}
\end{center}
\caption{PDF's of $N_{vis}(u)$ (\textcolor{blue}{{\bf blue}}) and $N_{e}(u)$ 
 (\textcolor{red}{{\bf red}}), along with a simple multifractal model  
 for the PDF of $N_{vis}(u)$ (\textcolor{paleblue}{{\bf pale blue}})
 and a fit by a discrete Gaussian distribution for the PDF of $N_e(u)$ 
 (\textcolor{palered}{{\bf pale red}}). {\it Inset:} The PDF of the 
 shift $\Delta N(u)$ in each realization. Standard errors of the mean for $N_{vis}(u)$ and $N_{e}(u)$ are smaller than the marker size.}
\lb{multi-pdf} \end{figure}


The fluctuating nature of the equilibration scale $N_e(u)$ has
important implications for the scale at which thermal effects appear in
statistical averages. We consider the velocity
structure functions, which we define as \be \|u_n\|_p=\langle
|u_n|^p\rangle^{1/p} \lb{p-norm} \ee with a $p$-th root. Here $\langle
\cdot \rangle$ denotes the steady-state average. Just as for the
deterministic Sabra model \cite{lvov998improved}, the noisy Sabra model
exhibits power-law decay of structure functions with wavenumber, $ \|
u_n \|_p \propto k_n^{-\sigma(p)}$, where $\sigma(p)=\zeta(p)/p$ in
terms of the standard exponents $\zeta(p)$, as shown in Fig.~\ref{structfun}.
To the accuracy of our calculation, the corresponding exponents
$\sigma(p)$ for noisy and deterministic models are indistinguishable.

The inertial range intermittency is typically associated with nonlinear
dependence of scaling exponents $\sigma(p)$ on $p$. Looking at
structure functions with varying $p$ is also helpful for elucidation of
the interplay of intermittency and noise. Higher positive values of $p$
assign a higher weight to events with larger value of $|u_n|$. In
contrast negative values of $p$ shift weight to events with smaller
values of $|u_n|$. This effect is clearly seen in Fig. \ref{structfun},
where one can read off the cut-off shellnumber $N_e(p)$ at which the
$p$th structure-function attains its equipartition value
$[\Gamma(1+\frac{p}{2})]^{1/p}\theta_K$.  Evidently $N_e(p)$ exhibits
strong $p$-dependence, with smaller $N_e(p)$ for negative values of
$p$, and larger values of $N_e(p)$ for large positive $p$.
Significantly, for negative values of $p$, thermal effects can be seen
even at length scales above $\eta,$ inside the traditional inertial
range. This suggests that a similar result should hold for the fluctuating
Navier-Stokes equation, but experimental observation will be extremely
challenging since the small magnitude of the thermal velocities, 3-4
orders of magnitude less than the Kolmogorov velocity $u_\eta,$ will
require very high precision of velocity measurements to be resolved.

\textit{Effect of spatial dimension.} Up to now, we have ignored the
potentially profound effect of space dimensionality $d$ on the
character of fluctuating hydrodynamics. This effect can be simply
understood by the estimate $u_\ell^{th}\sim (k_BT/\rho \ell^d)^{1/2}$
for the magnitude of a thermal velocity fluctuation at length-scale
$\ell$ in space-dimension $d$. It follows that the Reynolds number of
the fluctuating hydrodynamic equation (\ref{FNS}) in the thermal
equipartition regime is $$ Re^{th}_\ell :=
\frac{u_\ell^{th}\ell}{\nu}=\left(\frac{k_BT}{\rho\nu^2\ell^{d-2}}\right)^{1/2}$$
at length-scale $\ell$. Clearly for $d>2,$ $Re_\ell^{th}$ becomes
increasing large with decreasing scale $\ell$. For $d<2$ $Re_\ell^{th}$
becomes smaller with decreasing $\ell$. This fact is well-known and can
be obtained by more sophisticated renormalization group analysis
\cite{forster1976long,forster1977large}, according to which thermal
fluids are UV asymptotically free for $d<2$ but UV strongly coupled for
$d>2.$ These considerations apply to our shell model which has
effectively $d=0$ and, indeed, we observe in our simulations that the
noisy turbulent model becomes almost UV free at high-wavenumbers, with
individual shell modes $u_n$ governed to good approximation by
independent linear Langevin equations. Since 3D fluctuating
hydrodynamics instead becomes strongly coupled in the UV, it might be
doubted that our shell model properly represents the physics at scales
where thermal fluctuations become important. However, the justification
for our shell model lies in the extremely small values of $u_\ell^{th}$
for turbulent flows at scales $\ell\lesssim \eta.$ A dimensionless measure 
of this smallness is $Re^{th}_\eta=\theta_K^{1/2},$ which is of order 
$10^{-4}$ in the atmospheric boundary layer.  A general calculation 
\cite{eyink2021dissipation} confirms that fluctuating hydrodynamics
for any molecular fluid in $d=3$ will remain weakly coupled until 
$\ell$ is comparable to the size of a molecule! Thus, the strong-coupling regime 
in the model at scales smaller than this is of no physical relevance.


\textit{Conclusion.} In this Letter we provided theoretical order of
magnitude estimates, corroborated by numerical simulations of the Sabra
shell model of the turbulent cascade that thermal noise cannot be
neglected in the dissipation range of incompressible turbulence, in
contrast to the naive expectation that thermal effects are only
relevant in molecular fluids at scales comparable to the mean free path
length \cite{vonneumann1963recent}. Our arguments and numerical results 
thus confirm and extend the prescient ideas of Betchov 
\cite{betchov1957fine,betchov1961thermal,betchov1964measure}. 
At the moment there are almost no experimental measurements below the
Kolmogorov length to challenge the naive expectation that thermal noise
becomes important only at the scale of the mean free path. However,
there are many practically relevant processes that take place at
sub-Kolmogorov scales of turbulent flow and therefore there is an
urgent need for new experimental methods that can probe velocity
fluctuations at those small scales with high accuracy to check the
predictions made here. Perhaps the first evidence of thermal effects
will be indirect, through the competing effects of turbulent and
thermal fluctuations on the various physical phenomena strongly
sensitive to sub-Kolmogorov turbulent scales. For example, we expect to
see the influence of thermal noise on the Batchelor range of high
Schmidt-number passive scalars, which extends well below the Kolmogorov
scale.
Another implication of our results is that there should be a possibly
subtle influence of thermal noise on the inertial range of turbulence,
through sensitivity of the equations of motion to perturbations \cite{mailybaev2016spontaneously, thalabard2020butterfly}.

This work was supported by the Simons Foundation through Grant
number 662985 (NG) and Grant number 663054 (GE).

\bibliography{bibliography.bib}
\end{document}